\documentclass[10pt]{article}
\usepackage[latin9]{inputenc}
\usepackage{amsmath}
\usepackage{amssymb}

\makeatletter

\usepackage{amsfonts}\@ifundefined{definecolor}
 {\usepackage{color}}{}
\newcommand{\comm}[1]{}

\newcommand{\bey}{\begin{eqnarray}}

\newcommand{\beq}{\begin{equation}}
\newcommand{\eeq}{\end{equation}}
\newcommand{\bea}{\begin{array}}
\newcommand{\eea}{\end{array}}

\makeatother

\begin{document}

\begin{titlepage} \vskip 2cm 

\begin{center}
\textbf{\Large Classic tests of General Relativity described by brane-based spherically symmetric solutions}\textbf{ }%
\footnote{\texttt{rodrigo.cuzinatto@unifal-mg.edu.br,\!\;pedropjp@ifi.cta.br,}
\texttt{mdemonti@ualberta.ca,} \texttt{fkhanna@ualberta.ca, (khannaf@uvic.ca),}
\texttt{hoff@feg.unesp.br}%
}\textbf{ }
\par\end{center}

\begin{center}
\textbf{\vskip 2cm R. R. Cuzinatto$^{a}$, P. J. Pompeia$^{b,c}$,
M. de Montigny$^{d,e}$,}\\
\textbf{ F. C. Khanna$^{d,f,g}$, J. M. Hoff da Silva$^{h}$}\\
 \vskip 5pt \textsl{$^{a}$Instituto de Ci\^encia e Tecnologia, Universidade
Federal de Alfenas,}\\
\textsl{ Rodovia Jos\'e Aur\'elio Vilela, 11999, Cidade Universit\'aria,
CEP 37715-400}\\
\textsl{ Po\c cos de Caldas, MG, Brazil}
\par\end{center}

\begin{center}
\vskip 2pt \textsl{$^{b}$ Instituto de Fomento e Coordena\c c\~ao Industrial,
Departamento de Ci\^encia e Tecnologia Aeroespacial,}\\
\textsl{ Pra\c ca Mal. Eduardo Gomes 50, 12228-901, S\~ao Jos\'e dos Campos,
SP, Brazil }\\
\vskip 2pt \textsl{$^{c}$ Instituto Tecnol\'ogico de Aeron\'autica,
Departamento de Ci\^encia e Tecnologia Aeroespacial,}\\
\textsl{ Pra\c ca Mal. Eduardo Gomes 50, 12228-900, S\~ao Jos\'e dos Campos,
SP, Brazil}
\par\end{center}

\begin{center}
\vskip 2pt \textsl{$^{d}$Theoretical Physics Institute, University
of Alberta,}\\
\textsl{ Edmonton, AB, Canada T6G 2J1}\\
 \vskip 2pt \textsl{$^{e}$Campus Saint-Jean, University of Alberta,
}\\
\textsl{ Edmonton, AB, Canada T6C 4G9}\\
 \vskip 2pt \textsl{$^{f}$TRIUMF, 4004, Westbrook Mall,}\\
\textsl{ Vancouver, BC, Canada V6T 2A3}\\
 \vskip 2pt \textsl{$^{g}$Dept. of Physics \& Astronomy,
  University of Victoria,}\\
\textsl{ PO box 1700, STN CSC, Victoria, BC, Canada V8W 2Y2}\\
 \vskip 2pt \textsl{$^{h}$Departamento de F\'isica e Qu\'{i}mica,
Universidade Estadual Paulista, }\\
\textsl{ Av. Ariberto Pereira da Cunha 333, 12.516-410, Guaratinguet\'a,
SP, Brazil}\\
 \vskip 2pt 
\par\end{center}

\vskip .5cm 
\begin{abstract}
We discuss a way to obtain information about higher dimensions from
observations by studying a brane-based spherically symmetric solution.
The three classic tests of General Relativity are analyzed in details:
the perihelion shift of the planet Mercury, the deflection of light
by the Sun, and the gravitational redshift of atomic spectral lines.
The braneworld version of these tests exhibits an additional parameter
$b$ related to the fifth-coordinate. This constant $b$ can be constrained
by comparison with observational data for massive and massless particles.
\end{abstract}
\noindent {\em Keywords:} Braneworld solutions; Classic tests of
General Relativity. \\
 {\em PACS: 04.20.-q, 04.20.-h, 12.10.-g} \end{titlepage}




\section{Introduction}

Braneworld models have attracted considerable attention of the scientific
community since the outstanding works by L. Randall and R. Sundrum
\cite{RSI,RSII}. The possibilities raised in such a framework have been
extensively explored since then. In fact, from particle physics to
cosmology, a plethora of braneworld models were investigated.
In particular, the idea of standard model fields living only on the
brane, a necessity in \cite{RSI}, was rapidly overcome \cite{UNI}.

At least from the gravitational point of view, the very idea behind
braneworld models rests upon our belief that at high enough energies, General Relativity (GR) shall be at least corrected.
In this vein, the new scenario provided by the braneworld picture
has served as an interesting framework to cosmologists \cite{COSMO}.
Again, this time within cosmology, from inflation to large scale consequences,
the new possibilities for phenomenology provided by the braneworld
paradigm have been extensively investigated (for a broad review, see \cite{MAAR}). An important point to be stressed,
however, is that even far below high energy scales (which points to
the transition between classical and quantum gravity) at Solar System
size, there are interesting gravitational effects whose eventual modifications
arising from the braneworld that can be compared with experiments.

The aim of this work is to explore classic tests of GR in a spherically symmetric four-dimensional solution
embedded into a five-dimensional space. We use the metric

\begin{equation}
ds^{2}=-\left(1-\frac{2m}{\bar{r}}\right)(d\bar{x}^{4})^{2}+\frac{d\bar{r}^{2}}{\left(1-\frac{2m}{\bar{r}}\right)}+\bar{r}^{2}d\bar{\theta}^{2}+\bar{r}^{2}\sin^{2}\bar{\theta}d\bar{\varphi}^{2}+(d\bar{x}^{5})^{2},\label{BS}
\end{equation}
 where $\bar{x}^{5}$ stands for the extra dimension. Let us make
a few remarks about this expression. As usual, this line element
is obtained by the Schwarzschild four-dimensional solution embedded
into the extra dimension in the sense that at each $\bar{x}^{5}$
fixed slice we have the standard spherically symmetric solution. Obviously,
this line element can be related to the so-called black-string
\cite{BLACK}. Nevertheless it should be stressed that the solution presented in Eq.(\ref{BS}) is not necessarily related to the black-string,
i.e., the four-dimensional spherically symmetric metric need not
be related to a black hole. In fact, here we shall set up the mass
parameter to be far below the value necessary for a black-hole
solution, and investigate how the embedding of the solution into the
extra dimensional scenario can be related to the classic tests performed in GR. 

Let us point out that the classical tests of General Relativity have been examined for various spherically symmetric static vacuum solutions of braneworld models in Ref.  \cite{bohmer}. Therein, the authors exploit the Gauss-Codazzi approach in order to find corrections of the GR results by embedding the brane into the bulk. This must be accomplished by means of the Israel-Darmois junction conditions, which are valid only for singular branes, that is, if the brane is infinitely thin. This can be observed from Eq. (7) in Ref.  \cite{bohmer}. The constraints are imposed over the terms  ($E_{\mu\nu}$ in Eq. (17) of  \cite{shiromizu}) of the 5-dimensional Weyl tensor that carry information about the gravitational field outside the brane. Thus the main difference between the present study and Ref. \cite{bohmer} is that they deal with singular branes whereas we consider non-singular, or thick, branes \cite{dzhunushaliev}. Throughout the paper, branes will be understood in the sense that they are not necessarily singular. Note also that in Section 4.1 of Ref.  \cite{bohmer}, the authors claim that they obtain a DMPR-type solution (see Ref. \cite{dadhich}), which is the simplest solution for a spherically symmetric vacuum solution; this, too, does not contradict our results, since their results were found for singular branes, whereas our solution is a five-dimensional non-singular, non-spherically symmetric brane solution (only the four-dimensional section of our solution is spherically symmetric).

As we shall find, an additional parameter related to the
extra dimension can be constrained by these tests in a quite compatible
way for both, massive and massless test particles cases. We shall emphasize that in the context of universal extra dimensions \cite{UNI}, the fields must not be trapped on the brane but, instead, they are allowed to travel along the hole bulk. However, the extra dimension experienced by the fields shall be small (not to contradict $1/r^2$ deviations of Newton's law \cite{expe} in the case of gravitational experiments, and some key collider experiments \cite{KKlein}). Therefore the experimental boundaries applied in this work are used in order to viabilize an universal extra dimension from the point of view of classic Solar System tests.

This paper is structured as follows: after expressing the brane-based
spherically symmetric solution in light-cone coordinates, predictions
of the solution are confronted with three classic tests
of GR in Section \ref{sec-Tests}. In Section \ref{sec-PerihelionShift},
we find that the solution describes the perihelion shift of Mercury
as does four-dimensional Schwarzschild solution. In Section \ref{sec-DeflectionLight},
we observe that the solution predicts the deflection of light rays
by massive bodies like our Sun. In Section \ref{sec-Redshift}, we
obtain a similar result for the gravitational shift of atomic spectral
lines. The results of these tests depend on an additional parameter
$b$ (see Eq.(\ref{ge-vi-b})), which is related to the fifth-coordinate. This constant $b$
can be constrained by comparing with observational data and in Section 2.3, the result is interpreted via the uncertainty relations along the extra dimension. Section \ref{sec-FinalRemarks} contains some concluding remarks.


\section{Classic tests \label{sec-Tests}}



We consider a light-cone type transformation
in spherical coordinates, 
\begin{equation}
r=\bar{r},\;\;\;\theta=\bar{\theta},\;\;\;\varphi=\bar{\varphi},\;\;\; x^{4}=\frac{\bar{x}^{4}+\bar{x}^{5}}{\sqrt{2}},\;\;\; x^{5}=\frac{\bar{x}^{4}-\bar{x}^{5}}{\sqrt{2}}.\label{light-cone transf}
\end{equation}
 We shall recast the brane-based spherically symmetric solution in an appropriate way that suit our purpose.Hence when expressed in these coordinates, Eq. (\ref{BS}) becomes 
\begin{equation}
ds^{2}=\frac{dr^{2}}{1-\frac{2m}{r}}+r^{2}d\theta^{2}+r^{2}\sin^{2}\theta d\varphi^{2}+\frac{m}{r}\left(dx^{4}\right)^{2}+\frac{m}{r}\left(dx^{5}\right)^{2}+2\left(-1+\frac{m}{r}\right)dx^{4}dx^{5},\label{BS light-cone}
\end{equation}
 This line element describes the invariant interval on the curved
manifold in which the motion of particles and light rays will take
place. In the next section, we shall study these aspects in order
to test the brane-based spherically symmetric solution.


\subsection{Planetary motion \label{sec-PerihelionShift}}

The motion of test particles is described by the geodesic equations,
\[
\frac{d^{2}x^{\mu}}{ds^{2}}+\Gamma_{\rho\nu}^{\mu}\frac{dx^{\rho}}{ds}\frac{dx^{\nu}}{ds}=0.
\]
 In order to find solutions,this equation is written as 
\begin{align}
 & \left.\frac{d^{2}r}{ds^{2}}-\frac{m}{r^{2}}\frac{1}{\left(1-\frac{2m}{r}\right)}\left(\frac{dr}{ds}\right)^{2}-\left(1-\frac{2m}{r}\right)r\left(\frac{d\theta}{ds}\right)^{2}\right.\nonumber \\
 & \left.-\left(1-\frac{2m}{r}\right)r\sin^{2}\theta\left(\frac{d\varphi}{ds}\right)^{2}+\frac{m}{2r^{2}}\left(1-\frac{2m}{r}\right)\left(\frac{dx^{4}}{ds}+\frac{dx^{5}}{ds}\right)^{2}=0\right.,\label{geo r}
\end{align}
\begin{equation}
\frac{d^{2}\theta}{ds^{2}}+\frac{2}{r}\frac{dr}{ds}\frac{d\theta}{ds}-\sin\theta\cos\theta\left(\frac{d\varphi}{ds}\right)^{2}=0,\label{geo theta}
\end{equation}
\begin{equation}
\frac{d^{2}\varphi}{ds^{2}}+\frac{2}{r}\frac{dr}{ds}\frac{d\varphi}{ds}+2\cot\theta\frac{d\theta}{ds}\frac{d\varphi}{ds}=0,\label{geo phi}
\end{equation}
\begin{equation}
\frac{d^{2}x^{4}}{ds^{2}}+\frac{m}{r^{2}}\frac{1}{\left(1-\frac{2m}{r}\right)}\frac{dr}{ds}\left(\frac{dx^{4}}{ds}+\frac{dx^{5}}{ds}\right)=0,\label{geo x4}
\end{equation}
\begin{equation}
\frac{d^{2}x^{5}}{ds^{2}}+\frac{m}{r^{2}}\frac{1}{\left(1-\frac{2m}{r}\right)}\frac{dr}{ds}\left(\frac{dx^{4}}{ds}+\frac{dx^{5}}{ds}\right)=0.\label{geo x5}
\end{equation}
 We show that the motion lies in a plane as it happens in classical
mechanics for a central force \cite{Adler}. With an appropriate orientation
of the axis, we can choose the initial conditions to be 
\[
\theta_{0}=\frac{\pi}{2},\qquad\left(\frac{d\theta}{ds}\right)_{0}=0,
\]
 for some initial value of $s$. This choice implies that the motion
of the test particle starts at the ecliptic plane with zero initial
azimuthal velocity. It means also that Eq. (\ref{geo theta}) gives
a zero initial elevation acceleration, $\left(\frac{d^{2}\theta}{ds^{2}}\right)_{0}=0$.
Thus, at an infinitesimal proper instant later, $\theta_{\Delta s}=\frac{\pi}{2}$
and $\left(\frac{d\theta}{ds}\right)_{\Delta s}=0$, and similarly
after another $\Delta s$, and so on. As a result, the motion is confined
to the plane $\theta=\frac{\pi}{2}$ .

The geodesic equations, Eqs. (\ref{geo r}) to (\ref{geo x5}), are
then simplified to:
\begin{align}
 & \left.\frac{d^{2}r}{ds^{2}}-\frac{m}{r^{2}}\frac{1}{\left(1-\frac{2m}{r}\right)}\left(\frac{dr}{ds}\right)^{2}-\left(1-\frac{2m}{r}\right)r\left(\frac{d\varphi}{ds}\right)^{2}+\right.\nonumber \\
 & \left.+\frac{1}{2}\frac{m}{r^{2}}\left(1-\frac{2m}{r}\right)\left(\frac{dx^{4}}{ds}+\frac{dx^{5}}{ds}\right)^{2}=0,\right.\label{ge-i}
\end{align}
 
\begin{equation}
\frac{d^{2}\varphi}{ds^{2}}+\frac{2}{r}\frac{dr}{ds}\frac{d\varphi}{ds}=0,\label{ge-ii}
\end{equation}
\begin{equation}
\frac{d^{2}x^{4}}{ds^{2}}+\frac{m}{r^{2}}\frac{1}{\left(1-\frac{2m}{r}\right)}\frac{dr}{ds}\left(\frac{dx^{4}}{ds}+\frac{dx^{5}}{ds}\right)=0,\label{ge-iii}
\end{equation}
\begin{equation}
\frac{d^{2}x^{5}}{ds^{2}}+\frac{m}{r^{2}}\frac{1}{\left(1-\frac{2m}{r}\right)}\frac{dr}{ds}\left(\frac{dx^{4}}{ds}+\frac{dx^{5}}{ds}\right)=0.\label{ge-iv}
\end{equation} If we multiply Eq. (\ref{ge-ii}) by $r^{2}$, we find 
\begin{equation}
r^{2}\frac{d\varphi}{ds}=h,\label{ge-v}
\end{equation}
 where $h$ is a constant related to the conserved angular momentum
of the particle. Similarly, by adding Eqs. (\ref{ge-iii}) and (\ref{ge-iv}),
and by multiplying the result by $\left(1-\frac{2m}{r}\right)$, we
are led to 
\begin{equation}
\frac{dx^{4}}{ds}+\frac{dx^{5}}{ds}=\frac{k}{\left(1-\frac{2m}{r}\right)},\label{ge-vi}
\end{equation}
 where $k$ is another constant. By subtracting Eq. (\ref{ge-iii})
from Eq. (\ref{ge-iv}), we find 
\begin{equation}
\frac{dx^{4}}{ds}-\frac{dx^{5}}{ds}=b,\label{ge-vi-b}
\end{equation}
 where $b$ is constant. From Eqs. (\ref{ge-vi}) and (\ref{ge-vi-b}),
we obtain 
\begin{equation}
\frac{dx^{4}}{ds}=\frac{1}{2}\left[\frac{k}{\left(1-\frac{2m}{r}\right)}+b\right],\label{dx4/ds}
\end{equation}
 and 
\begin{equation}
\frac{dx^{5}}{ds}=\frac{1}{2}\left[\frac{k}{\left(1-\frac{2m}{r}\right)}-b\right].\label{dx5/ds}
\end{equation}

In principle both constants $k$ and $b$ could be associated to the extra dimension. However, as will be seen, the constant $k$ does not contribute to the orbital equation obtained below.

By substituting Eqs. (\ref{ge-v}) and (\ref{ge-vi}) into Eq. (\ref{ge-i}),
it follows that 
\[
\frac{d^{2}r}{ds^{2}}-\frac{m}{r^{2}}\frac{1}{\left(1-\frac{2m}{r}\right)}\left(\frac{dr}{ds}\right)^{2}-\left(1-\frac{2m}{r}\right)\frac{h^{2}}{r^{3}}+\frac{1}{2}\frac{m}{r^{2}}\frac{k^{2}}{\left(1-\frac{2m}{r}\right)}=0.
\]
 As in the classic Kepler problem, we can simplify the integration
processes by considering $r$ as a function of $\varphi$ instead
of $s$. If we change the variable $r$ to 
\[
u=\frac{1}{r},
\]
 it is possible to rewrite the last differential equation as 
\begin{equation}
\frac{d^{2}u}{d\varphi^{2}}+\frac{m}{\left(1-2mu\right)}\left(\frac{du}{d\varphi}\right)^{2}+\left(1-2mu\right)u-\frac{1}{2}\frac{k^{2}}{h^{2}}\frac{m}{\left(1-2mu\right)}=0.\label{ge-vii}
\end{equation}
 The term proportional to $\left(\frac{du}{d\varphi}\right)^{2}$
can be expressed in another form. For this, we use the constraint
\begin{equation}
g_{\mu\nu}\frac{dx^{\mu}}{ds}\frac{dx^{\nu}}{ds}=-1,\label{constraint particle}
\end{equation}
 which leads to 
\begin{equation}
\left(\frac{du}{d\varphi}\right)^{2}+\left(1-2mu\right)\left[u^{2}+\frac{1}{h^{2}}\left(1+\frac{b^{2}}{2}\right)\right]-\frac{1}{2}\frac{k^{2}}{h^{2}}=0.\label{ge-viii}
\end{equation}
 By inserting Eq. (\ref{ge-viii}) into Eq. (\ref{ge-vii}), we find
the orbital equation, 
\begin{equation}
\frac{d^{2}u}{d\varphi^{2}}+u-\frac{m}{h^{2}}-3mu^{2}-\frac{m}{h^{2}}\frac{b^{2}}{2}=0.\label{ge-ix}
\end{equation}
 The first four terms are the usual ones obtained in the standard
4-dimensional GR. The extra term should provide corrections related
to the additional dimension.

Let us make an important remark concerning large distances. If this
case is considered then the terms proportional to $u^{2}$ in Eq.
(\ref{ge-ix}) should be neglected, thus
\begin{equation}
\frac{d^{2}u}{d\varphi^{2}}+u-\frac{m}{h^{2}}-\frac{m}{h^{2}}\frac{b^{2}}{2}=0\,.\label{Newton orbit}
\end{equation}
 The first three terms are the usual terms obtained by Newtonian gravitation.
The additional term, proportional to $b^{2}$, is open for interpretation.
If agreement with measurements are to be obtained, then either $b$
should be negligibly small (which would leave us with the usual GR
result) or $b$ should be included in a renormalized value of $m$.
In Section \ref{sec-DeflectionLight} we shall use $b\ll1$.


\subsubsection{The perihelion shift}

Let us rewrite Eq. (\ref{ge-ix}) as follows: 
\begin{equation}
\frac{d^{2}u}{d\varphi^{2}}+u=\frac{m}{\bar{h}^{2}}+3mu^{2},\label{orbit eq}
\end{equation}
 with
\begin{equation}
\frac{1}{\bar{h}^{2}}=\frac{1}{h^{2}}\left(1+\frac{b^{2}}{2}\right).\label{h bar}
\end{equation}
 Formally the orbital equation is exactly as predicted
by GR, the only difference being the redefinition $h\rightarrow\bar{h}$.
Then, we know beforehand that the solution in question will predict
a perihelion shift for the orbit of the planets consistent with GR.

The usual procedure is to obtain a solution of Eq. (\ref{orbit eq})
through an iterative procedure taking $u\simeq u^{\left(0\right)}+u^{\left(1\right)}$
\cite{Sabbata}. The zero-order is the unperturbed solution of Eq.
(\ref{Newton orbit}), or Eq. (\ref{orbit eq}) with $3mu^{2}=0$.
It is utilized as a source term for the differential equation of $u^{\left(1\right)}$,
i.e. we shall write $3mu^{2}=3m\left(u^{\left(0\right)}\right)^{2}$.
The integration constants are the eccentricity of the orbit $e$ and
an arbitrary initial value $\varphi_{0}$ for the azimuthal angle.
The constant $e$ is related to the major axis $a=r_{\max}$ by 
\begin{equation}
a=\frac{L}{1-e^{2}},\label{a}
\end{equation}
 where
\begin{equation}
L=\frac{m}{\bar{h}^{2}}\label{semilatus}
\end{equation}
 is the semi-latus rectum of the orbit. We proceed as in GR by considering orbits of small
eccentricity (like the ones of Mercury)and find that the perihelion
shifts after a full revolution by
\begin{equation}
\Delta\varphi_{0}=6\pi\frac{m^{2}}{\bar{h}^{2}}=6\pi\frac{m}{L}=6\pi\frac{GM}{c^{2}a\left(1-e^{2}\right)}.\label{2pi shift}
\end{equation}


\subsubsection{Numerical analysis for the planet Mercury}

Here, the quantities of interest are expressed in terms of orbital
parameters of the planet under consideration and the geometrical mass
of the Sun. Let us consider the planet Mercury, with the following
orbital data \cite{Seidelmann}: 
\begin{align*}
a & =0.38709893\, AU=57.909\,175\times10^{6}\, \rm{km},\\
e & =0.20563069.
\end{align*}
 The numerical value of the geometrical mass \cite{PDG} of the Sun is: 
\begin{equation}
m=\frac{GM_{\mathrm{Sun}}}{c^{2}}=1.4766250385\left(1\right)\, \rm{km}.\label{m}
\end{equation} Hence the value of $L$ for Mercury is
\[
L=\left(1-e^{2}\right)a=55.460545\times10^{6}\, \rm{km}
\]
and the perihelion shift, from Eq. (\ref{2pi shift}), is 
\[
\Delta\varphi_{0}=1.59748705\pi\times10^{-7}=0.10351716{}^{\prime\prime}.
\]
 This is an extremely small angle, but this is a secular effect which
increases with the number of revolutions. The shift above is observed
in a single Mercury-year; which corresponds to $0.24084960$ Earth-years
\cite{Seidelmann}. So, the total shift per Earth-year is
\[
\Delta\varphi_{E}=\frac{\Delta\varphi_{0}}{0.24084960}=0.4298{}^{\prime\prime}.
\]
 If this effect is accumulated over 100 Earth-years, the total shift
is 
\[
\Delta\varphi=100~\Delta\varphi_{E}=42.98{}^{\prime\prime}.
\]

The conclusion is that the solution is quite similar to the one obtained with GR Schwarzschild
solution for a prediction of the perihelion shift of Mercury. The
difference is that we can calculate the value for the constant $\bar{h}$\ \ while
in GR we obtain directly the value of $h$. The relative difference
that would be obtained using GR calculations and the one done here
is
\begin{eqnarray*}
\left|\frac{\Delta\varphi_{GR}-\Delta\varphi}{\Delta\varphi_{GR}}\right| & = & \left|1-\frac{1}{\left(1+\frac{b^{2}}{2}\right)}\right|=\frac{b^{2}}{2}\frac{1}{\left(1+\frac{b^{2}}{2}\right)}.
\end{eqnarray*}
If we consider that ``the excess shift is known to about 0.1 percent''
\cite{CWill}, then this difference can be used to evaluate an upper
limit for the values of $b$ for Mercury. In this case, $\left|b\right|<0.045$.
Of course this analyzis does not take into account the parametrized post-newtonian (PPN) corrections.
If this was done then the upper limit for $b$ would certainly be
smaller.

In 1997, Tegmark argued that there exist no stable orbit in a four-dimensional spacetime \cite{tegmark}, giving rise to a stability problem for the spacetime described by Eq. (\ref{BS}). Let us remark that the assertion of instability in the $(4,1)$ case is based on Ref. \cite{tangherlini}, whose analysis is performed on an $n$-dimensional spherically symmetric line element (see Eq. (3.1) of Ref. \cite{tangherlini}). Our Eq. (\ref{BS}) is spherically symmetric only in four dimensions. Within the context of non-singular branes, the fields are localized around the brane core, but not restricted to a four-dimensional slice of the spacetime. The extra dimension being small (that is, the fields being restricted to a small part of the extra dimension), there is no problem with the motion through the bulk and no stability problems. If the fifth dimension (or the fourth {\it space-like} dimension) were infinite, then we would face instability problems. In our case, we have a three-dimensional spherical elements plus a {\it finite} fourth dimension. Then, rather than a $1/r^2$ potential (which would be the case in a four-dimensional manifold with all coordinates with an infinite domain), as mentioned in the last paragraph of Ref. \cite{tegmark}, we obtain a Yukawa-type potential, which allows stable orbits (see Section 3.3 of Ref. \cite{nozari}).


\subsection{The deflection of light rays \label{sec-DeflectionLight}}

Having investigated the parameter $b$ for massive particles, let us turn ourselves to the massless case. Light rays consist of massless test particles which travel at the
speed of light. In special relativity, the photons move along the
light-cone, following a null geodesic: $ds^{2}=0$. We will keep $ds^{2}=0$ for photon traveling in our background. So the framework is a curved manifold described by the brane-based spherically
symmetric solution on which the relativistic particles will
propagate. We feel justified in doing so because the photons are test
particles and, by definition, test particles do not affect the geometry
of the background spacetime.

By taking $ds^{2}=0$ in the Schwarzschild-like spacetime, Eq. (\ref{BS light-cone}),
and dividing the result by $d\sigma^{2}$\ (where $\sigma$ is an
appropriate invariant length), the line element becomes
\begin{eqnarray}
 &  & \left.\frac{1}{\left(1-\frac{2m}{r}\right)}\left(\frac{dr}{d\sigma}\right)^{2}+r^{2}\left(\frac{d\theta}{d\sigma}\right)^{2}+r^{2}\sin^{2}\theta\left(\frac{d\varphi}{d\sigma}\right)^{2}\right.\nonumber \\
 & + & \left.\frac{m}{r}\left(\frac{dx^{4}}{d\sigma}+\frac{dx^{5}}{d\sigma}\right)^{2}-2\frac{dx^{4}}{d\sigma}\frac{dx^{5}}{d\sigma}=0.\right.\label{constraint light}
\end{eqnarray}
 This constraint replaces the one given by Eq. (\ref{constraint particle})
for massive particles. All the equations before Eq. (\ref{constraint particle})
remain valid for the propagation of light, provided that we replace
the invariant length $s$ by $\sigma$; that is, 
\begin{align}
 & \left.\frac{d^{2}r}{d\sigma^{2}}-\frac{m}{r^{2}}\frac{1}{\left(1-\frac{2m}{r}\right)}\left(\frac{dr}{d\sigma}\right)^{2}-\left(1-\frac{2m}{r}\right)r\sin^{2}\theta\left(\frac{d\varphi}{d\sigma}\right)^{2}+\right.\nonumber \\
 & \left.\qquad\quad+\frac{1}{2}\frac{m}{r^{2}}\left(1-\frac{2m}{r}\right)\left(\frac{dx^{4}}{d\sigma}+\frac{dx^{5}}{d\sigma}\right)^{2}=0,\right.\label{ge-i light}
\end{align}
\begin{equation}
\frac{d\varphi}{d\sigma}=\frac{h}{r^{2}},\label{ge-v light}
\end{equation}
\begin{equation}
\frac{dx^{4}}{d\sigma}=\frac{1}{2}\left[\frac{k}{\left(1-\frac{2m}{r}\right)}+b\right],\label{dx4/ds light}
\end{equation}
\begin{equation}
\frac{dx^{5}}{d\sigma}=\frac{1}{2}\left[\frac{k}{\left(1-\frac{2m}{r}\right)}-b\right],\label{dx5/ds light}
\end{equation}
 where the initial conditions are $\theta_{0}=\pi/2$\ and $\left(\frac{d\theta}{d\sigma}\right)_{0}=0$.
These conditions imply $\left(\frac{d^{2}\theta}{d\sigma^{2}}\right)_{0}=0$
and restrict our study to the plane $\theta=\pi/2$. Therefore, Eq.
(\ref{ge-vii}) is still valid for the light rays. However, Eq. (\ref{ge-viii})
must be modified, since it was obtained using $ds^{2}\neq0$, $g_{\mu\nu}u^{\mu}u^{\nu}=-1$,
whereas here we have $ds^{2}=0$, $g_{\mu\nu}u^{\mu}u^{\nu}=0$.

Let us rewrite the new constraint, Eq. (\ref{constraint light}),
by substituting $\theta=\pi/2$, $d\theta/d\sigma=0$, together with
Eqs. (\ref{ge-v light}), (\ref{dx4/ds light}) and (\ref{dx5/ds light}):
\[
\left(\frac{dr}{d\sigma}\right)^{2}+\left(1-\frac{2m}{r}\right)\left(\frac{b^{2}}{2}+\frac{h^{2}}{r^{2}}\right)-\frac{k^{2}}{2}=0.
\]
 Since 
\[
\frac{dr}{d\sigma}=\frac{dr}{d\varphi}\frac{d\varphi}{d\sigma}=\frac{dr}{d\varphi}\frac{h}{r^{2}},
\]
 we have 
\[
\left(\frac{dr}{d\varphi}\right)^{2}+\left[\left(1-\frac{2m}{r}\right)\left(\frac{b^{2}}{2}+\frac{h^{2}}{r^{2}}\right)-\frac{k^{2}}{2}\right]\frac{r^{4}}{h^{2}}=0.
\]
 Now this equation is written as a function of $u=1/r$:
\begin{equation}
\left(\frac{du}{d\varphi}\right)^{2}+\left(1-2mu\right)\left[u^{2}+\frac{1}{h^{2}}\frac{b^{2}}{2}\right]-\frac{1}{2}\frac{k^{2}}{h^{2}}=0,\label{ge-viii light}
\end{equation}
 which differs only slightly from our previous constraint, Eq. (\ref{ge-viii}). By substituting Eq. (\ref{ge-vii}) into Eq. (\ref{ge-viii light}),
one obtains 
\begin{equation}
\frac{d^{2}u}{d\varphi^{2}}+u-3mu^{2}+\frac{m}{h^{2}}\frac{b^{2}}{2}=0.\label{light eq b}
\end{equation}

If $b\simeq0$, we observe that Eq. (\ref{light eq b}) reduces to
the equation obtained in the standard 4-dimensional GR leading to
the deflection of light rays. In analogy with the previous subsection, we first set $3mu^{2}=0$,
in order to get an approximate solution of Eq. (\ref{light eq b})
by an iterative procedure, starting with a zero-order solution, 
\begin{equation}
u^{\left(0\right)}=\frac{1}{R}\cos\left(\varphi-\varphi_{0}\right)-\frac{m}{h^{2}}\frac{b^{2}}{2},\label{u(0) light}
\end{equation}
 where $\varphi_{0}$\ and $R$\ are integration constants. The
interpretation of $R$ becomes clear when we set $\varphi_{0}=0$,\ and
introduce a Cartesian coordinate system $x=r\cos\varphi$, $y=r\sin\varphi$,\ with
origin $O$ at the center of the massive body, which is the source
of the field.

With $b=0$, we find that Eq. (\ref{u(0) light}) reduces to $R=r\cos\varphi=x$.
This is a straight line parallel to the $y$-axis, and $R$ is the
minimum distance between the light ray and the origin $O$. In this
case, $u^{\left(0\right)}$ does not bend the straight trajectory
of the photon and there is no deflection of light. However, this is
not the best possible approximation, and there is also a $b\neq0$ contribution to be taken into account.

The first approximation to Eq. (\ref{light eq b}) leads to 
\[
\frac{d^{2}u^{\left(1\right)}}{d\varphi^{2}}+u^{\left(1\right)}=-\frac{m}{h^{2}}\frac{b^{2}}{2}+3m\left(u^{(0)}\right)^{2},
\]
 which becomes
\begin{align*}
\frac{d^{2}u^{\left(1\right)}}{d\varphi^{2}}+u^{\left(1\right)} & =-\frac{m}{h^{2}}\frac{b^{2}}{2}+\frac{3m}{R^{2}}\cos^{2}\varphi\\
 & +\frac{3m^{3}}{h^{4}}\left(\frac{b^{4}}{4}\right)-6\frac{m^{2}}{h^{2}}\left(\frac{b^{2}}{2}\right)\frac{1}{R}\cos\varphi.
\end{align*}
 It is to be noted that if we consider $b\ll1$, then the factor of $b^{4}$ can be neglected, and we get
\[
\frac{d^{2}u^{\left(1\right)}}{d\varphi^{2}}+u^{\left(1\right)}=\frac{3m}{R^{2}}\cos^{2}\varphi-\frac{m}{h^{2}}\frac{b^{2}}{2}-6m\frac{m}{h^{2}}\frac{b^{2}}{2}\frac{1}{R}\cos\varphi.
\]
 A particular solution of this differential equation is 
\[
u^{\left(1\right)}=\frac{m}{R^{2}}\left(\cos^{2}\varphi+2\sin^{2}\varphi\right)-\frac{m}{h^{2}}\frac{b^{2}}{2}-3m\frac{m}{h^{2}}\frac{b^{2}}{2}\frac{1}{R}\varphi\sin\varphi.
\]
 Therefore, 
\begin{align}
u & \simeq u^{\left(0\right)}+u^{\left(1\right)},\nonumber \\
 & =\frac{1}{R}\cos\varphi+\frac{m}{R^{2}}\left(\cos^{2}\varphi+2\sin^{2}\varphi\right)-\frac{m}{h^{2}}\frac{b^{2}}{2}\left(2+\frac{3m}{R}\varphi\sin\varphi\right).\label{u light}
\end{align}
 Notice that if $b=0$ this equation reduces to the expression derived
in GR \cite{Sabbata}. This also means that all possible modifications
predicted by the braneworld picture for the deflection of light are present in the last term of Eq. (\ref{u light}).

If we multiply Eq. (\ref{u light}) by $rR$, we have 
\[
R=r\cos\varphi+\frac{m}{R}\left(r\cos^{2}\varphi+2r\sin^{2}\varphi\right)-\frac{m}{h^{2}}\frac{b^{2}}{2}\left[\left(2R\right)r+\left(3m\varphi\right)r\sin\varphi\right].
\]
 Then, in Cartesian coordinates, we have 
\[
x=R-\frac{m}{R}\left(\frac{x^{2}+2y^{2}}{\sqrt{x^{2}+y^{2}}}\right)+\frac{m}{h^{2}}\frac{b^{2}}{2}\left[2R\sqrt{x^{2}+y^{2}}+\left(3m\arctan\frac{y}{x}\right)y\right].
\]
 The second term on the r.h.s gives the GR's deviation of the light
ray from the straight line $x=R$. The last term is the contribution
arising from the extra dimension. In the limit where $y\gg x$ (which means great distances from the source), we obtain the asymptotic solution: 
\[
x\simeq R-\frac{m}{R}\left(\pm2y\right)+\frac{m}{h^{2}}\frac{b^{2}}{2}\left[2R\left(\pm y\right)+\left(3m\frac{\pi}{2}\right)y\right],
\]
 since $\lim_{\alpha\rightarrow+\infty}\left(\arctan\alpha\right)=\pi/2$.
Thus, the two possible values of $x$ are 
\[
x_{+}=R+\frac{2m}{R}y+\frac{m}{h^{2}}\frac{b^{2}}{2}\left(2R+\frac{3\pi}{2}m\right)y,
\]

and 

\[
x_{-}=R-\frac{2m}{R}y+\frac{m}{h^{2}}\frac{b^{2}}{2}\left(-2R+\frac{3\pi}{2}m\right)y,
\]
 and the deflection is described by the angle
\[
\tan\delta\simeq\frac{x_{+}-x_{-}}{y}.
\]
 With $\tan\delta=\delta+O\left(\delta^{3}\right)$, the deflection
angle is 
\begin{equation}
\delta=\frac{4m}{R}+\left(2mR\right)\frac{b^{2}}{h^{2}}=\frac{m}{R}\left(4+2\frac{R^{2}b^{2}}{h^{2}}\right).\label{deflection}
\end{equation}

A comparison with experimental data \cite{Shapiro2004},
where the deflection for the case under consideration would be
\[
\delta=\frac{m}{R}\left(3.99966\pm0.00090\right),
\]
shows that $2\frac{R^{2}b^{2}}{h^{2}}$ is constrained to be $2\frac{R^{2}b^{2}}{h^{2}}<0.00056\Rightarrow\left|\frac{b}{h}\right|<\sqrt{\frac{0.00056}{2R^{2}}}$.
Using the value of radius of the Sun \cite{SolarRadius},
$R=\left(696,342\pm65\right)\rm{km}$, we find$\left|\frac{b}{h}\right|<2.403015\times10^{-8}\rm{km}^{-1}$.


\subsection{Gravitational redshift of spectral lines \label{sec-Redshift}}

Next we examine the shift of the atomic spectral lines
in the presence of a gravitational field, also called the \emph{gravitational
redshift}.

From the line element (\ref{BS light-cone}) we define 
\[
d\tau^{2}=-\frac{1}{2}ds^{2}
\]
 as the time-interval between two events with vanishing spatial separation,
$dr=d\theta=d\varphi=0$. The minus sign is a result of our choice for the signature of the metric, and the factor $1/2$ is chosen in order
to allow agreement with the four-dimensional analog (Schwarzschild
solution) when $b=0$.

The time-interval is related to the coordinate time differential,
$dx^{4}$, and the additional brane differential coordinate, $dx^{5}$,
by 
\begin{equation}
-2d\tau^{2}=\frac{m}{r}\left(dx^{4}\right)^{2}+\frac{m}{r}\left(dx^{5}\right)^{2}+2\left(-1+\frac{m}{r}\right)dx^{4}dx^{5},\label{d tau metric}
\end{equation}
 where the differentials are constrained by Eqs. (\ref{dx4/ds}) and
(\ref{dx5/ds}): 
\[
\frac{dx^{4}}{d\tau}=\frac{1}{2}\left[\frac{k}{\left(1-\frac{2m}{r}\right)}+b\right],\qquad\frac{dx^{5}}{d\tau}=\frac{1}{2}\left[\frac{k}{\left(1-\frac{2m}{r}\right)}-b\right],
\]
 so that
\begin{equation}
dx^{5}=dx^{4}-bd\tau.\label{dx5(b)}
\end{equation}
 It is to be noted that these are general equations; they are valid
for massive particles, but they have exactly the same form as for
massless particles, such as photons.

Now the time-interval in Eq. (\ref{d tau metric})
is expressed as 
\begin{equation}
d\tau^{2}=\left(1-\frac{2m}{r}\right)\left(dx^{4}\right)^{2}-\frac{m}{r}\frac{b^{2}}{2}d\tau^{2}-\left(1-\frac{2m}{r}\right)bdx^{4}d\tau.\label{time int}
\end{equation}
 For the special case of $b=0$, we have 
\begin{equation}
d\tau=+\sqrt{1-\frac{2m}{r}}dx^{4},\label{d tau b=00003D0}
\end{equation}
which is the expected Schwarzschild solution of GR (see Ref. \cite{Sabbata},
Eq. (4.92)). The positive sign follows from the natural assumption
that the proper time $\tau$ should increase with the time coordinate
$x^{4}$. With $b\neq0$, Eq. (\ref{time int}) leads to 
\[
\left(1+\frac{m}{r}\frac{b^{2}}{2}\right)d\tau^{2}+\left(1-\frac{2m}{r}\right)b~dx^{4}d\tau-\left(1-\frac{2m}{r}\right)\left(dx^{4}\right)^{2}=0.
\]
 This is a second-order equation for $d\tau$. By solving for $d\tau=d\tau\left(dx^{4}\right)$,
we find 
\[
d\tau=\frac{-\left(1-\frac{2m}{r}\right)b~dx^{4}\pm\sqrt{\left(1-\frac{2m}{r}\right)^{2}b^{2}\left(dx^{4}\right)^{2}+4\left(1+\frac{m}{r}\frac{b^{2}}{2}\right)\left(1-\frac{2m}{r}\right)\left(dx^{4}\right)^{2}}}{2\left(1+\frac{m}{r}\frac{b^{2}}{2}\right)}.
\]
 In order that this expression agrees with Eq. (\ref{d tau b=00003D0})
for $b=0$, we must choose the plus sign in the r.h.s., which leads
to 
\begin{equation}
d\tau=\left[\frac{\sqrt{1+\frac{b^{2}}{4}}-\frac{b}{2}\sqrt{1-\frac{2m}{r}}}{\left(1+\frac{2m}{r}\frac{b^{2}}{4}\right)}\right]\sqrt{1-\frac{2m}{r}}dx^{4}.\label{d tau}
\end{equation}

This distinction between proper time and the time coordinate gives
rise to a difference between the proper frequency and the coordinate
frequency of a periodic phenomenon in the curved spacetime, like the
emission of electromagnetic radiation by an atom. Consider the propagation
of electromagnetic waves in the limit of geometrical optics. Then,
the electromagnetic field can be written as $f=a\exp\left(i\psi\right)$,
with $a=a\left(\mathbf{x},t\right)$ the wave amplitude, and the phase
$\psi=\psi\left(\mathbf{x},t\right)$ an eikonal function. Then the
frequency of the wave can be expressed as the derivative of $\psi$
with respect to the time, and one has a coordinate frequency, $\omega_{0}=\frac{\partial\psi}{\partial t}$,
and a proper frequency, $\omega=\frac{\partial\psi}{\partial\tau}$.

We call $\omega_{1}$ the proper frequency of the wave emitted by
an atom at a point $P_{1}$. At another point, $P_{2}$, the observed
proper frequency will be different, say $\omega_{2}$, once the gravitational
field is not the same. They are related so that 
\[
\frac{\omega_{2}}{\omega_{1}}=\frac{\left(\frac{\partial\psi}{\partial\tau}\right)_{2}}{\left(\frac{\partial\psi}{\partial\tau}\right)_{1}}=\frac{\frac{\partial\psi}{\partial t}\left(\frac{\partial t}{\partial\tau}\right)_{2}}{\frac{\partial\psi}{\partial t}\left(\frac{\partial t}{\partial\tau}\right)_{1}}=\frac{\omega_{0}\left(\frac{\partial x^{4}}{\partial\tau}\right)_{2}}{\omega_{0}\left(\frac{\partial x^{4}}{\partial\tau}\right)_{1}},
\]
 with $\omega_{0}$ constant and where $x^{4}$ is the time coordinate
$t$. From Eq. (\ref{d tau}), we find 
\[
\left(\frac{dx^{4}}{d\tau}\right)_{1}=\frac{\left(1+\frac{2m}{r_{1}}\frac{b^{2}}{4}\right)}{\sqrt{1+\frac{b^{2}}{4}}-\frac{b}{2}\sqrt{1-\frac{2m}{r_{1}}}}\frac{1}{\sqrt{1-\frac{2m}{r_{1}}}}
\]
 and similarly for $\left(dx^{4}/d\tau\right)_{2}$. Then, the ratio
of frequencies is 
\begin{equation}
\frac{\omega_{2}}{\omega_{1}}=\frac{\left(\frac{\partial x^{4}}{\partial\tau}\right)_{2}}{\left(\frac{\partial x^{4}}{\partial\tau}\right)_{1}}=\frac{\left(1+\frac{2m}{r_{2}}\frac{b^{2}}{4}\right)}{\left(1+\frac{2m}{r_{1}}\frac{b^{2}}{4}\right)}\frac{\left(\sqrt{1+\frac{b^{2}}{4}}-\frac{b}{2}\sqrt{1-\frac{2m}{r_{1}}}\right)}{\left(\sqrt{1+\frac{b^{2}}{4}}-\frac{b}{2}\sqrt{1-\frac{2m}{r_{2}}}\right)}\frac{\sqrt{1-\frac{2m}{r_{1}}}}{\sqrt{1-\frac{2m}{r_{2}}}}.\label{w2/w1}
\end{equation}

This exact result can be approximated by taking into account the fact
that regular estimates assume $r_{1}\ll r_{2}$, and that $b$ is
small.\ For
instance, let us assume that the radiation observed on Earth, at $r=r_{2}\approx 149.6\times10^{6} \, \rm{km}$, was emitted at the surface of the Sun, at $r=r_{1}=R\approx 696 \, \rm{km}$. Then the approximation $r_{1}\ll r_{2}$ is certainly valid. Moreover,
the quantity $m/r_{1}$\ is very small, since $m=1.4766250385\left(1\right)\, \rm{km} $, and
the functions in Eq. (\ref{w2/w1})\ can be approximated accordingly.
Therefore, for $m/r_{1}\ll1$\ and $m/r_{2}\ll1$, the ratio of frequencies
is 
\[
\frac{\omega_{2}}{\omega_{1}}\simeq\left(1+\frac{2m}{r_{2}}\frac{b^{2}}{4}\right)\left(1-\frac{2m}{r_{1}}\frac{b^{2}}{4}\right)\frac{\left[\sqrt{1+\frac{b^{2}}{4}}-\frac{b}{2}\left(1-\frac{m}{r_{1}}\right)\right]}{\left[\sqrt{1+\frac{b^{2}}{4}}-\frac{b}{2}\left(1-\frac{m}{r_{2}}\right)\right]}\left(1-\frac{m}{r_{1}}\right)\left(1+\frac{m}{r_{2}}\right).
\]
 If m/r2  is assumed near zero, then this ratio simplifies to
\[
\frac{\omega_{2}}{\omega_{1}}\simeq\left(1-\frac{2m}{r_{1}}\frac{b^{2}}{4}\right)\frac{\left[\sqrt{1+\frac{b^{2}}{4}}-\frac{b}{2}\left(1-\frac{m}{r_{1}}\right)\right]}{\left[\sqrt{1+\frac{b^{2}}{4}}-\frac{b}{2}\right]}\left(1-\frac{m}{r_{1}}\right).
\]
 Furthermore, if $b$ is also assumed to be very small, $b<<1$, this
equation becomes 
\begin{equation}
\frac{\omega_{2}}{\omega_{1}}\simeq1-\frac{m}{r_{1}}\left(1-\frac{b}{2}\right),\label{w2/w1 approx}
\end{equation}
 so that 
\[
\frac{\Delta\omega}{\omega}=\frac{\omega_{2}-\omega_{1}}{\omega_{1}}=-\frac{m}{R}\left(1-\frac{b}{2}\right).
\]
 When $b=0$, this result agrees with GR. The value predicted by GR
using the solar geometrical mass and radius used earlier leads to a
redshift of $\left|\frac{\Delta\omega}{\omega}\right|=2.120546\times10^{-6}$.
In velocity scale, i.e. converting this redshift as if it was due
to a doppler effect, this redshift corresponds to $v\approx636\, \rm{m\, s^{-1}}$.
The value estimated in Ref.\cite{SolarRedshift} is $633\, \rm{m\, s^{-1}}$
with an uncertainty $\lesssim100\, \rm{m\, s^{-1}}$. If we consider this value we can fix an upper limit for $\left|b\right|$ using the
lower experimental value $533\, \rm{m\, s^{-1}}$. We obtain $\left|b\right|<0.323$.
If instead of considering the lower limit we had taken the estimated
value we would have obtained $\left|b\right|<0.00857$. This would
be the case if the uncertainty of measurements can be considerably
reduced and if the estimated value remains the same.

Now if reconsider the case of light deflection and take this former
value as a limit for $\left|b\right|$ for light rays then we are
able to estimate for the values of 
\[
\left|\frac{b}{h}\right|<2.403015\times10^{-8}{\rm{km}}^{-1}<\frac{0.323}{\left|{h}\right|}\Rightarrow\left|{h}\right|<1.34\times10^{8}\rm{km}.
\]

We conclude this section by interpreting the obtained upper limit to the $b$ parameter in terms of the fifth dimension by using the uncertainty relations. As one can see, the $b$ parameter can be directly related to the extra dimension. Therefore, its upper limit also means a limit on the velocity along $x^{5}$.   

Let us assume for the argument the dispersion relation for a photon in a 4-dimensional
space:
\[
p^{2}=p_{\mu}p^{\mu}=0\Rightarrow E^{2}=\mathbf{p}^{2}c^{2}\Rightarrow E=\left|\mathbf{p}\right|c=h\nu,
\]
where $E$ is the energy, $\mathbf{p}$ is the 3-momentum, $c$ is
the velocity of light, $h$ is the Planck constant and $\nu$ is the
frequency of the photon.

If we assume a quantum behavior for the photon, then the following
uncertainty relations are expected \cite{Birula}:
\begin{eqnarray*}
\Delta E\Delta t & \gtrsim & \frac{\hbar}{2}\\
\Delta x_{i}\Delta p_{i} & \gtrsim & \frac{\hbar}{2},\; i=1,2,3.
\end{eqnarray*} Hence, if the fifth dimension is supposed to exist, we would expect the uncertainty relations to be generalized as
\begin{eqnarray*}
\Delta E\Delta t & \gtrsim & \frac{\hbar}{2}\\
\Delta x_{i}\Delta p_{i} & \gtrsim & \frac{\hbar}{2},\; i=1,2,3,5,
\end{eqnarray*}
and the dispersion relation as
\begin{eqnarray*}
p^{2} & = & 0\Rightarrow E^{2}=\left(\mathbf{p}^{2}+p_{5}^{2}\right)c^{2}\Rightarrow E^{2}=\mathbf{p}^{2}c^{2}\left(1+\frac{p_{5}^{2}}{\mathbf{p}^{2}}\right)\Rightarrow\\
 &  & \Rightarrow E=\left|\mathbf{p}\right|c\sqrt{1+\frac{p_{5}^{2}}{\mathbf{p}^{2}}}.
\end{eqnarray*}
If $\frac{\left|p_{5}\right|}{\left|\mathbf{p}\right|}\ll1$, then
we can approximate the relation as 
\[
E\approx\left|\mathbf{p}\right|c\left(1+\frac{1}{2}\frac{p_{5}^{2}}{\mathbf{p}^{2}}\right).
\]
When comparing this result with the 4-dimensional case, it is possible to interpret
the extra term as a fluctuation of energy (which is supposed to be
small since all influences of the fifth-dimensional quantities are
not directly observed):
\[
\Delta E\approx\frac{1}{2}c\frac{p_{5}^{2}}{\left|\mathbf{p}\right|}.
\]
If this fluctuation of energy is supposed to be consistent with uncertainty
relations, then we would expect, with the best accuracy:
\[
\Delta E\Delta t\approx\frac{\hbar}{2}\Rightarrow\frac{1}{2}c\frac{p_{5}^{2}}{\left|\mathbf{p}\right|}\Delta t\approx\frac{\hbar}{2}.
\]
During this time interval $\Delta t$, the distance traveled
by light in the fifth dimension is 
\[
\Delta x^{5}=bc\Delta t.
\]
 If $p_{5}\approx\Delta p_{5}$, which means that the magnitude of $p_{5}$ is comparable to its uncertainty, then we obtain
\[
\frac{1}{2}\frac{\left(\Delta p_{5}\right)^{2}}{\left|\mathbf{p}\right|}\frac{\Delta x^{5}}{b}\approx\frac{\hbar}{2}\Rightarrow\frac{1}{2}\frac{\left(\Delta p_{5}\right)^{2}}{\left|\mathbf{p}\right|b}\left(\Delta x^{5}\right)^{2}\approx\frac{\hbar}{2}\Delta x^{5}.
\]
As stated earlier, if we suppose that the accuracy of measurements is the best
possible, then $\Delta p_{5}\Delta x^{5}\approx\frac{\hbar}{2}$,
so that
\[
\frac{1}{2}\left(\frac{\hbar}{2}\right)^{2}\frac{1}{\left|\mathbf{p}\right|b}\approx\frac{\hbar}{2}\Delta x^{5}\Rightarrow\frac{1}{8\pi}\frac{1}{b}\frac{h}{\left|\mathbf{p}\right|}\approx\Delta x^{5}.
\]
From the de Broglie relation, $\frac{h}{\left|\mathbf{p}\right|}=\lambda$,
we find
\[
\frac{1}{8\pi}\frac{1}{b}\lambda\approx\Delta x^{5}.
\]

If we assume that $\Delta x^{5}$ represents an estimate
of the size of the fifth dimension, and keeping in mind that we have obtained
an upper limit for $b$, then we conclude that our results lead to:
\[
b<b_{0}\Rightarrow\Delta x^{5}\gtrsim\frac{1}{8\pi}\frac{1}{b_{0}}\lambda.
\]
In our case, the value of $b_{0}$ is estimated with results from experiments
that used a wavelength in the range of 5188-5212$\textrm{\AA}$. We
use the average value to estimate $\Delta x^{5}$ as follows:
\[
\Delta x^{5}\gtrsim\frac{1}{8\pi}\frac{1}{b_{0}}\lambda=\frac{1}{8\pi}\frac{1}{0.323}5200\times10^{-10}\ {\mathrm m}=6.4\times10^{-8}\ {\mathrm m}.
\]

Therefore, we find that the uncertainty on the measured position along the fifth dimension is larger than the above value. The point to be stressed, however, is that this argument can now be read backwards: the very existence of the fifth dimension, when scrutinized by quantum particles, has an associated uncertainty related to its size implying an upper limit in the velocity, here parameterized by $b$.  

Let us note that if we consider $\Delta x^5$ as the uncertainty of measurement of lengths along the fifth-dimension and if the size of the fifth-dimension is smaller than the value estimated above, then we will not be able to distinguish its existence because there will not be sufficient accuracy for that. If the size of the fifth dimension is larger than shown by our calculations, then it would be possible to notice it. Therefore, no matter the size of the fifth dimension, our calculation just  established our capability of measuring it. In the absence of any other evidence, then we can conclude that the size of the fifth-dimension must be smaller than what we have estimated.



\section{Concluding remarks \label{sec-FinalRemarks}}

In this paper, we have investigated the embedding of a spherically
symmetric gravitational solution into the five-dimensional braneworld
scenario. The classic tests of GR are considered in order to study
the possible influence of the extra dimension in low energy experiments. 
In fact, we have examined three GR classic tests: the perihelion shift
of Mercury, the deflection of light by the Sun, and the gravitational
redshift of atomic spectral lines. The investigated solution gives
results similar to the 4D Schwarzschild line element predictions.
More precisely, a new parameter is related to the extra dimension that brings subtle but important corrections to the usual GR case can be constrained in order not to contradict any experimental results. Moreover, at least in the massless (photon) case, it is possible to interpret the obtained results with fundamental concepts as the uncertainty principle.

Finally we wish to stress the relevance of the aforementioned analysis. Firstly it can be used to refine the extra dimensional models of General Relativity, by constraining the new parameters using
experimental observations. In addition the present study can help in explaining small differences observed in gravitational experiments. The exploration of GR classical tests,with the possible departures, may serve as an important tool in connecting high energy models to low energy experiments.

\bigskip{}

\section*{Acknowledgement}

We acknowledge partial support by the Natural Sciences and Engineering
Research Council (NSERC) of Canada. RRC and PJP acknowledge financial
support from FAPEMIG (Brazil) and CNPq (Brazil). JMHS acknowledge
financial support from CNPq. The authors are grateful to Prof. V. P. Frolov, Drs. A. Zelnikov, H. Yoshino, D. Gorbonos and A. Shoom for useful comments and fruitful discussions. We thank the anonymous referee for many insightful comments that helped us to improve our manuscript.

\bigskip{}


\begin{thebibliography}{10}
\bibitem{RSI} L. Randall and R. Sundrum, Phys. Rev. Lett. \textbf{83},
3370 (1999).

\bibitem{RSII} L. Randall and R. Sundrum, Phys. Rev. Lett. \textbf{83},
4690 (1999).

\bibitem{UNI} T. Appelquist, H.-C. Cheng, and B. A. Dobrescu, Phys. Rev.
\textbf{D 64}, 035002 (2001).

\bibitem{COSMO} P. Binetruy, C. Deffayet, U. Ellwanger, and D. Langlois,
Phys. Lett. \textbf{B 477}, 285 (2000); P. Bowcock, C. Charmousis, and 
R. Gregory, Class. Quantum Grav. \textbf{17}, 4745 (2000); A. Campos and C. F. Sopuerta, Phys. Rev. \textbf{D 63}, 104012 (2001); A. Campos and C. F. Sopuerta, Phys. Rev. \textbf{D 64}, 104011 (2001); S. Mukohyama,
T. Shiromizu, and K. Maeda, Phys. Rev. \textbf{D 62}, 024028 (2000).

\bibitem{MAAR} R. Maartens and K. Koyama, ``Brane-World Gravity'', Living Rev. Relativity \textbf{13}, 5 (2010). URL
(accessed February 06nd, 2014): http://relativity.livingreviews.org/Articles/lrr-2010-5/.

\bibitem{BLACK} A. Chamblin, S. W. Hawking, and H. S. Reall, Phys. Rev.
\textbf{D 61}, 065007 (2000).

\bibitem{bohmer} C.G. B\"ohmer, G. De Risi, T. Harko, and F. S. N. Lobo, Class. Quant. Grav. \textbf{27}, 185013 (2010).

\bibitem{shiromizu} T. Shiromizu, K. Maeda, and M. Sasaki, Phys. Rev. D \textbf{62}, 024012 (2000)

\bibitem{dzhunushaliev} V. Dzhunushaliev, V. Folomeev, and M. Minamitsuji, Rep. Prog. Phys. \textbf{73}, 066901 (2010).

\bibitem{dadhich} N. Dadhich, R. Marteens, P. Papadopoulos, and V. Rezania, Phys. Lett. B \textbf{487}, 1 (2000).

\bibitem{expe}  D. J. Kapner, T. S. Cook, E. G. Adelberger, J. H. Gundlach, B. R. Heckel, C. D. Hoyle, and H. E. Swanson, Phys. Rev. Lett. \textbf{98}, 021101 (2007). 

\bibitem{KKlein} M. Baak, M. Goebel, J. Haller, A. Hoecker, D. Ludwig, K. Moenig, M. Schott, and J. Stelzer, Eur. Phys. J. \textbf{C 72}, 2003 (2012).

\bibitem{Adler}R. Adler, M. Bazin, and M. Schiffer, {Introduction
to General Relativity}, 2nd edition, MacGraw-Hill Book Co., New York,
(1975).


\bibitem{Seidelmann}K. P. Seidelmann (Ed.) - Explanatory Supplement
to the Astronomical Almanac, University Science Books, Mill Valley,
California (1992).

\bibitem{PDG}J. Beringer et al. (Particle Data Group), Phys. Rev.
\textbf{D 86}, 010001 (2012).

\bibitem{CWill}C. M. Will, ``The Confrontation between General Relativity
and Experiment'' , Living Rev. Relativity \textbf{9}, 3 (2006). URL
(accessed January 22nd, 2014): http://www.livingreviews.org/lrr-2006-3. 


\bibitem{tegmark} M. Tegmark, Class. Quant. Grav. \textbf{14}, L69 (1997)

\bibitem{tangherlini} F. R. Tangherlini, Nuov. Cim. \textbf{27}, 636 (1963)

\bibitem{nozari} K. Nozari and S. Akhshabi, Chaos, Solitons and Fractals, \textbf{37}, 324 (2008)

\bibitem{Sabbata}V. de Sabbata and M. Gasperini, {Introduction to
Gravitation}, World Scientific Publishing Co., Singapore, (1985).

\bibitem{Shapiro2004}S. S. Shapiro,\emph{ et al}, Phys. Rev. Lett.
\textbf{92}, 121101 (2004).

\bibitem{SolarRadius}M. Emilio, \emph{et al,} Astrophys. J. \textbf{750},
135 (2012).

\bibitem{SolarRedshift}Y. Takeda and S. Ueno, Solar Phys \textbf{281},
55 (2012).

\bibitem{Birula} I. Bialynicki-Birula and Z. Bialynicki-Birula, Phys. Rev. Let. \textbf{108}, 140401 (2012). \end{thebibliography}
\end{document}